\documentclass[9pt]{article}
\usepackage{subeqnarray,IEEEtrantools}
\usepackage[title]{appendix}
\usepackage{epsfig,amsmath,amssymb,mathtools,dirtytalk,csquotes}
\usepackage{enumitem,setspace}
\usepackage{mathrsfs}
\usepackage{empheq}
\usepackage{tikz,lipsum,lmodern}
\usepackage[most]{tcolorbox}

\usepackage{soul,ulem,authblk}
\usepackage{centernot}
\usepackage[pagebackref=true, colorlinks=true]{hyperref}
\definecolor{redish}{rgb}{0.7,0.2,0.0}  
\definecolor{bluish}{rgb}{0.2,0.5,0.8}
\hypersetup{linkcolor=redish,          
	citecolor=blue,        
	filecolor=magenta,      
	urlcolor=bluish}          

\def \({\left(}
\def \){\right)}
\def \[{\left[}
\def \]{\right]}

\begin{document}
\title{Unprovability of First Maxwell's Equation in Light of EPR's Completeness Condition -- A Computational Approach from Logico-linguistic Perspective}
\author{Abhishek Majhi\footnote{abhishek.majhi@gmail.com}}%
\affil{{\small Indian Statistical Institute,\\Plot No. 203, Barrackpore,  Trunk Road,\\ Baranagar, Kolkata 700108, West Bengal, India}}
\date{~}
\maketitle 
\begin{abstract}
 Maxwell's verbal statement of Coulomb's experimental verification of his hypothesis, concerning force between two electrified bodies, is suggestive of a modification of the respective computable expression on logical grounds. This modification is in tandem with the completeness condition for a physical theory, that was stated by Einstein, Podolsky and Rosen in their seminal work. Working with such a modification, I show that the first Maxwell's equation, symbolically identifiable as ``$\vec{\nabla}\cdot\vec{E}=\rho/\epsilon_0$'' from the standard literature, is {\it unprovable}. This renders Poynting's theorem to be {\it unprovable} as well. Therefore, the explanation of `light' as `propagation of electromagnetic energy' comes into question on theoretical grounds. 
 \\\vspace{0.1cm}

 {\it Keywords:}  Coulomb's hypothesis; EPR Completeness Condition; Maxwell's equations; Mathematical Logic; Choice; Decision problem.
\end{abstract}

\section{Introduction}
Nobody has seen light, I believe. Rather light lets us `see'. Thus, propagation of light is a hypothesized phenomenon which is theoretically established by the equations which were written by Maxwell\cite{maxwelleq}, and what we know today as Maxwell's equations\cite{jackson, griffiths}. Today we have the understanding of light propagation as ``propagation of electromagnetic wave'', which forms the basis of our explanations of reflection, refraction, interference, diffraction, polarization and other optical phenomena\cite{bornwolf}. That the propagation of light is indeed  ``propagation of electromagnetic energy'' is theoretically established by a theorem which was proved by Poynting\cite{poynting}, and what we know today as Poynting's theorem\cite{jackson, griffiths}. {\color{black}Based on such theoretical establishment, the intuition of light propagation is transferred from classical to quantum regime leading to the hypothesis of traveling (unseen) `photons' e.g. see refs.\cite{smith2007,aspect2007,qoptics2008,centphoton} and the relevant references therein.}
However, Einstein's special theory of relativity suggests that the one way propagation speed of light is not knowable within the theory because the theory itself is founded on a choice of clock synchronization convention that is based on the assumption of a two-way velocity of light, which Einstein elaborated before stating the two postulates\cite{einsr}. Such a choice can be safely  considered as an independent axiom of special relativity and it has continued to remain a matter of  investigation and debate since the advent of special relativity\cite{stanconvensim}. Therefore, if there is any claim that Maxwell's equations are consistent with the special relativity, then the two way velocity of light is also a necessary assumption for the relativistic Maxwell's equations, which indeed was the case for Einstein's own work\cite{einsr}. 
This leaves room for skepticism regarding the mathematical structure that provides us the foundation, and hence the logic, behind our confidence on the theoretical propagation model given by Maxwell's equations. Such skepticism seems justified from Poincare's bold declaration from the stand point of mathematical reasoning, on page no. 6 of ref.\cite{poincare}, that ``{\it it is precisely in the proofs of the most elementary theorems that the authors of classic treatises have displayed the least precision and rigour.}'' It was such precision and rigour, from the mathematical logical perspective, that Hilbert searched for through his sixth problem \cite{hilbertprob}.

The motto of the present work is to investigate the proof of the first Maxwell's equation, symbolically identifiable as ``$\vec{\nabla}\cdot\vec{E}=\rho/\epsilon_0$'' from standard modern textbooks\cite{griffiths,jackson} and also known as the differential form of Gauss' law, by considering the steps of reasoning in propositional form so that mathematical logic becomes vividly applicable\cite{hilbertacker,mendelson,kleene1}. This is a particular, nevertheless a significant, attempt to implement Hilbert's  philosophy that is manifested in his sixth problem, namely, ``{\it Mathematical Treatment of the Axioms of Physics}''\cite{hilbertprob,wm1} or ``axiomatization of physics'' in modern terminology\cite{gorban}. As I expose through a propositional truth analysis, the first Maxwell's equation is written by halting at a decision problem, created by choice, so as to make room for a further suitable choice, by defying the rules of logic, to write the desired result. Therefore, the first Maxwell's equation is formally unprovable; it can only be written by {\it choice}\cite{turing1}. Since the standard ``proof'' of Poynting's theorem\cite{poynting}, as available in the modern standard textbooks\cite{jackson,griffiths}, is dependent on the first Maxwell's equation, therefore Poynting's theorem becomes formally unprovable as a consequence. The preciseness of the computable expressions corresponding to the verbal statements of Coulomb's hypothesis (`law') stated by  Maxwell in his book\cite{maxwell}, in light of the completeness condition of a physical theory stated by Einstein, Podolsky and Rosen\cite{epr}{\color{black}, or EPR Completeness Condition (ECC) in short},  plays the pivotal role in this analysis. {\color{black}{To mention, by ``computable expression'' I mean an expression with which one can perform both mathematical and logical computations.}}

{\color{black} I may note that it is indeed unusual to discuss classical physics in light of ECC, especially in the modern era of quantum science\cite{nobel2022} where the word ``EPR'' gets uniquely associated with quantum physics e.g. see refs.\cite{bellstan,eprency} and the references therein. However, the present analysis concerning the basics of classical electrodynamics only brings out a broader aspect of ECC that underlies the logic and language of physics in general, irrespective of `classical' and `quantum' distinction. In particular, there are two aspects of this work which are worth noting. Firstly, the loss of theoretical ground behind our confidence in the propagation model of light as waves, provides enough ground to rethink about the hypothesis of `traveling photons along particular paths'(e.g. see refs.\cite{smith2007, aspect2007, qoptics2008, centphoton} and the references therein) that lays the foundation of experiments involving entangled photons\cite{nobel2022}. This can potentially open up new ways of thinking about the metaphysics of such experiments\cite{expmeta} and reignite debates concerning the law of causation in physics e.g. see refs.\cite{causestan,wikicausality} and the relevant references therein. Secondly, the present analysis lays the ground for a quest to reformulate the theory of interaction between two charged bodies with a provision of explaining Maxwell's statement of Coulomb's hypothesis through computation (see  Appendix(\ref{appA}) for a glimpse of such a possibility). 	Such a quest concerns the language of physics and its computational content -- a slowly but steadily emerging research direction motivated by foundational questions\cite{ll1,lpp,chupqg,majhidot,gisin1,gisin2}.
		
}

{\color{black} Let me debrief the structure of the article as follows. In section(\ref{sec2}), at first I revisit Coulomb's hypothesis as was stated by Maxwell in his book {\it The Treatise of Electricity and Magnetism: Vol. 1}\cite{maxwell} and modify the respective computable expression to seek a more precise and truthful conversion of the verbal expressions. I explain how such modification is a necessity in light of ECC\cite{epr}.  In section (\ref{sec3}), I rewrite the definition of electric field due to a source point charge in symbolic terms, keeping track of the associated logical conditions, considering the modified computable expression of Coulomb's hypothesis as the premise and compute the divergence to showcase the unprovability of the first Maxwell's equation, albeit informally. 
In section (\ref{sec4}), I revisit some basic facts and tools of logic and cast the Dirac delta function in terms of logical connectives, along with explanations through truth tables. In section (\ref{sec5}),  I discuss formally, in terms of logical connectives, how the process of human reasoning leads to the first Maxwell's equation through a sequence of choices, of which, the first one creates a decision problem by halting the computation, that provides the room for a second  choice to suitably write the desired equation. I discuss the process of reasoning in detail for a source point charge and consequently generalize it for a source configuration of non-overlapping point charges and a continuous distribution of charge. Thus, I expose the formal unprovability of the first Maxwell's equation and, hence, the Poynting's theorem. In section (\ref{sec6}), I conclude with some remarks. }

  In accord with the standard practice in the modern literature\cite{griffiths,jackson}, I use Dirac delta function\cite{dirac} (henceforth, to be called as delta function) to present my analysis. Also, I mean ``classical logic'' while using the word ``logic'' in this work, following the standard practice in mathematical logic e.g. see ref.\cite{hilbertacker,mendelson,kleene1}. 

\section{Maxwell's statement of Coulomb's hypothesis and EPR completeness condition}\label{sec2}
Before analyzing Maxwell's statement of Coulomb's hypothesis, let me provide the motivation to indulge in such an inquiry, which will automatically set the tone of the rest of this work and lay down the attitude with which this work needs to be studied. 

While explaining Einstein's operational analysis of concepts that formed the basis of special relativity\cite{einsr}, Bridgman pointed out, on page no. 5 of ref.\cite{logicphysics}, that a ``{\it concept is synonymous with the corresponding set of operations.}''  As concepts are expressed through language, while writing the theories of science, there comes the question of the truthfulness of expressions of experience (operations) in terms of verbal language. Then follows the question of how precisely the verbal language is encoded in the respective computable expressions. So, it boils down to how precise one can make the following interrelations: 
\begin{eqnarray}
	\text{physical operations}\leftrightarrow \text{expressions through verbal language}\leftrightarrow\text{computable expressions }.\nonumber 
\end{eqnarray}
It is the second interrelation that is under investigation in the present discussion. 

The importance of language and associated reasoning in physics has been manifested earlier\cite{hilbertprob,wm1} and also it has  recently gained emphasis from various independent research investigations\cite{gorban, ll1, lpp, chupqg, gisin1, gisin2}. Maintaining the attitude of making logico-linguistic, or semantically driven, inquiry into the foundations of physics that has been showcased in ref.\cite{ll1,chupqg}, in what follows, I demonstrate an important consequence of a truthful conversion of verbal statement to the respective computable expression that concerns Coulomb's hypothesis regarding the force between two charged or electrified bodies.

On p 69 of ref.\cite{maxwell}, Maxwell wrote the following:\vspace{0.1cm}

``{\it Coulomb showed by experiment that the force between electrified bodies whose dimensions are small compared with the distance between them, varies inversely as the square of the distance. Hence the actual repulsion between two such bodies charged with quantities $e$ and $e^{\prime}$ and placed at a distance $r$ is }
\begin{eqnarray}
	\frac{ee^{\prime}}{r^2}.\text{''}\label{maxcl}
\end{eqnarray}

Here, the matter of inquiry is whether the computable expression, that is (\ref{maxcl}), actually encodes what Maxwell stated verbally in the preceding statements to explicate the set of physical operations that Coulomb needed to perform to experimentally verify his hypothesis or `law'. The significance of such an inquiry becomes manifest through  EPR's Completeness Condition (ECC) for a ``physical theory''\cite{epr}:\vspace{0.1cm}

 ``{\it every element of the physical reality must have a counterpart in the physical theory}''.\vspace{0.1cm}

 I explain as follows. Maxwell's verbal statement about the comparison of dimensions and the distance is an expression of the experience of the experimenter (Coulomb) in the laboratory. Such ``sense experiences'' of the experimenter is what I consider as ``physical reality'', following Einstein\cite{einphreal}, which is expressed in terms of language to inscribe a physical theory. So, according to ECC, if the computable expression (\ref{maxcl}) is a part of a physical theory, then it should encode the verbal statement about the comparison between ``dimensions'' and ``distance''. Now, ``dimensions'' and ``distance'' can be compared if and only if ``dimensions'' means ``length dimensions''; otherwise the  comparison is meaningless due to mismatch of physical dimensions, which is nevertheless a basic lesson of metrology\cite{vim} and, in particular, of dimensional analysis\cite{dimana}. So, certainly there should be an association of some characteristic lengths with the electrified bodies, say, $s$ and $s^{\prime}$. Then, the computable expression (\ref{maxcl}), being the building block of a physical theory, should be accompanied by the condition ``$s, s^{\prime}< r$''. Therefore, in order to satisfy ECC, (\ref{maxcl}) should be rewritten as follows: 
\begin{eqnarray}
	\frac{ee^{\prime}}{r^2}:s, s^{\prime}< r,\label{maxclss}
\end{eqnarray}  
where the symbol ``$:$'' stands for ``such that''.

{\color{black}Now}, the standard practice is to work with the hypothetical notion of ``point charge''\cite{jackson}. One can object that as far as physical reality is concerned (i.e. in light of ECC), nobody has ever experienced ``a point''\cite{majhidot}, and hence,  ``a point charge''. It is only to implement the axiom of point, from geometry, that the notion of ``a point'' is invoked. While I admit that this is a valid objection that should be taken into account, I may also assert that the consequence would be a drastic change in the computable expressions of electrostatics, which I plan to discuss elsewhere e.g. see Appendix (\ref{appA}) for a glimpse and an estimate of $s, s'$.  For the present purpose, I choose to work with ``point charge'' in accord with the standard convention, which is necessary for investigating the proof of the first Maxwell's equation as it appears in the standard literature. Therefore, ignoring any hint of doubt regarding whether such a practice is feasible or not from the ECC point of view, I consider the following modification of Maxwell's verbal statement in terms of ``point charge'':\vspace{0.1cm} 

{\it Coulomb showed by experiment that the force between electrified bodies, each considered as a point charge, not overlapping with each other, or equivalently, having a non-vanishing distance between them, varies inversely as the square of the distance.} \vspace{0.1cm}

Such a verbal statement can be encoded in the computable expression by considering the condition ``$r\neq 0$'' to be mandatory.  On such grounds of reasoning, the expression (\ref{maxcl}) should now be modified, in the standard jargon of ``point charge'', as follows:
\begin{eqnarray}
\frac{ee^{\prime}}{r^2}:r\neq 0.\label{maxclpc}
\end{eqnarray}  
Now, I write the content of (\ref{maxclpc}) in a more general form by introducing a coordinate system so as to write it in terms of vector notation following the modern textbook standard practice\cite{griffiths,jackson}. 

I consider a test charge $q_0$ at field point $\vec{r}$ and a source charge $q_i$ at source point $\vec{r}_i$, where I consider both the charges to be positive definite (as per general convention). According to Coulomb's hypothesis, the force on the test charge $q_0$, due to the source charge $q_i$ situated at $\vec{r}_i$, is written as follows:
\begin{eqnarray}
	\vec{F}_{(q_i,\vec{r}_i)}(\vec{r})=\frac{q_0q_i}{4\pi\epsilon_0}\frac{(\vec{r}-\vec r_i)}{|\vec r-\vec r_i|^3}\quad:\vec{r}\neq\vec r_i\label{coulombecc}
\end{eqnarray}
I may note that (\ref{coulombecc}) is the expression of Coulomb's hypothesis, in modern standard notation, that takes into account the ECC. Now, in what follows, I define the notion of electric field based on (\ref{coulombecc}) and do the respective computations.

\section{Electric field due to a source point charge {\color{black} and its divergence}}\label{sec3} 
Considering (\ref{coulombecc}) as the premise, I define the electric field at the field point $\vec{r}$,  due to $q_i$ situated at $\vec{r}_i$, as follows:
\begin{eqnarray}
\vec{E}_{(q_i, \vec{r}_i)}(\vec{r})&:=& \lim_{q_0\to 0}\frac{\vec{F}_{(q_i, \vec{r}_i)}}{q_0}=\lim_{q_0\to 0}\frac{q_0}{q_0}\cdot\frac{q_i}{4\pi\epsilon_0}\frac{(\vec{r}-\vec r_i)}{|\vec r-\vec r_i|^3}~:\vec{r}\neq\vec r_i   \nonumber\\
&=& \frac{q_i}{4\pi\epsilon_0}\frac{(\vec{r}-\vec r_i)}{|\vec r-\vec r_i|^3}~:\vec{r}\neq\vec r_i.~~~~[\because \lim_{q_0\to 0}\frac{q_0}{q_0}=1]\label{epc}
\end{eqnarray}
This definition is in accord with the standard practice except the use of symbols so as to have a more precise correspondence with the verbal language. Therefore, I may write the divergence of ``$\vec{E}_{(q_i, \vec{r}_i)}(\vec{r})$'' as follows:
\begin{eqnarray}
	\vec{\nabla}\cdot\vec{E}_{(q_i, \vec{r}_i)}(\vec{r})&=& \frac{q_i}{4\pi\epsilon_0}\vec{\nabla}\cdot\frac{(\vec{r}-\vec r_i)}{|\vec r-\vec r_i|^3}~:\vec{r}\neq\vec r_i\nonumber\\
&=& \frac{q_i}{\epsilon_0}\delta^3(\vec{r}-\vec r_i)~:\vec{r}\neq\vec r_i\quad\left[\text{using}~~\vec{\nabla}\cdot\frac{(\vec{r}-\vec r_i)}{|\vec r-\vec r_i|^3}=4\pi\delta^3(\vec{r}-\vec r_i) \right]\label{divEdelta}.
\end{eqnarray}
 {\color{black} At this point, from (\ref{divEdelta}), it is now trivial to write:
	\begin{eqnarray}
		\vec{\nabla}\cdot\vec{E}_{(q_i, \vec{r}_i)}(\vec{r})
		&=& 0 ~:\vec{r}\neq\vec r_i~~\left[\because \delta^3(\vec{r}-\vec r_i)=0:\vec{r}\neq\vec r_i \right]\label{unprov}.
	\end{eqnarray}
There is no logical way, without invoking any choice, to write any statement for $\vec{r}=\vec{r}_i$. Hence, the unprovability (in light of ECC) of the first Maxwell's equation for a source point charge, which appears in the standard literature as ``$\vec{\nabla}\cdot\vec{E}_{(q_i, \vec{r}_i)}(\vec{r})
= \frac{q_i}{\epsilon_0}\delta^3(\vec{r}-\vec r_i)$'', already seems evident. However, then the following question arises -- how is the first Maxwell's equation written?  I intend to expose that the human process of reasoning leads to the first Maxwell's equation through choices -- at first a decision problem is created by a choice of halting the computation, by defying the rules of logic, and then a suitable choice is made to write the desired equation. I may note that decision problems are part of human reasoning, in general, based on which equations of physics are constructed e.g. see Appendix (\ref{appB}) for a discussion on an elementary decision problem that forms the basis of the continuity equation.

Such choices should not be  confused with gauge choice in electrodynamics. Rather, such choices can be likened to the choices made by an external operator of the choice machine of Turing\cite{turing1}, albeit with the following difference. In Turing's scenario a choice is made by the external operator when the machine halts at a decision problem. But, here the decision problem is created by choice so that further choices can be made to get the desired result.  To carry out such an investigation and demonstrate such choices, in the next section, I review some basis facts and tools of logic which I am going to use for the rest of the analysis. 
} 

{\color{black}
	\section{Basics of logic and writing the delta function in terms of logical connectives}\label{sec4}
	Here is the list of symbols for the logical connectives that I am going to use for further analysis: 
	\begin{itemize}
		\item  ``$\veebar$'' stands for ``logical exclusive disjunction''	(XOR)
		\item  ``$\vee$'' stands for ``logical inclusive disjunction'' (OR)
		\item  ``$\wedge$'' stands for ``logical conjunction'' (AND)
		\item  ``$\neg$'' stands for ``logical negation'' (NOT) 
		\item  ``$\rightarrow$'' stands for ``logical forward implication'' (IF with a forward sense; implies)
		\item  ``$\leftarrow$'' stands for ``logical reverse implication'' (IF with a reverse sense; implied by)
		\item  ``$\equiv$'' stands for ``logical equivalence'' (IFF; two way implication)
	\end{itemize}
I may note that I have used the adjective ``logical'' in each of the above cases so as to distinguish the use of those words in a formal sense (object language) from the use of those words in an informal sense (metalanguage) and to understand the use of quotation marks, one may consult Section 18 of ref.\cite{tarski}. The essence of the logical connectives can be understood from the following truth table for two arbitrary propositions $A$ and $B$:
	$$\begin{array}{|c|c|c|c|c|c|c|c|}\hline A & B & A\wedge B& A\vee B& A\veebar B&A\rightarrow B&A\leftarrow B&A\equiv B \\\hline F & F & F&F&F&T&T&T\\T&F&F&T&T&F&T&F \\F & T & F&T&T&T&F&F \\T& T & T&T&F&T&T &T\\\hline\end{array}$$
	I may note from the above truth table that $(A\equiv B)\equiv ((A\rightarrow B)\wedge(A\leftarrow B))$.
	
	Now, having introduced the symbols of the logical connectives, it is essential to revisit the three basic laws of logic. The three laws, for any arbitrary proposition $P$, are the following\cite{kleene1}:
	\begin{itemize}
	\item Law of Double Negation (LDN):	$\neg\neg P\equiv P$.
	\item Law of Excluded Middle (LEM): $P\vee\neg P$. 
	\item Law of Non-Contradiction (LNC): $\neg(P\wedge\neg P)$.
	\end{itemize} 
	The simultaneous truth of the above three laws is equivalent to $P\veebar\neg P$ i.e. ``$P\veebar\neg P$'' represents the Laws of Classical Logic (LCL). This becomes evident from observing that the $T$-s in the third and the sixth column correspond to the $T$-s of the last column of the following truth table. 
	$$\begin{array}{|c|c|c|c|c|c|c|}\hline P & \neg P &\overbrace{\neg\neg P\equiv P}^{\text{\footnotesize LDN}}& \overbrace{P\vee \neg P}^{\text{\footnotesize LEM}}& \overbrace{\neg(P\wedge\neg P)}^{\text{\footnotesize LNC}}& \text{\footnotesize[LEM]}\wedge\text{\footnotesize[LNC]}\wedge\text{\footnotesize[LDN]}& P\veebar\neg P  \\\hline 
		T & T & F & T & F & F & F \\
		F & F & F & F & T & F & F \\
		T & F & T & T & T & T & T \\
		F & T & T & T & T & T & T \\
		\hline\end{array}$$
	Thus, only the third and the fourth rows define the rules of logic. Having provided such clarifications, I proceed to investigate the role of choice in human reasoning to write the first Maxwell's equation. 	
}

{\color{black}
	The first step is to write the delta function using logical connectives. To begin with, I consider the following:
	\begin{itemize}
	\item ``$P$'' symbolizes ``$\vec{r}=\vec{r}_i$'',
	\item ``$\neg P$'' symbolizes ``$\vec{r}\neq\vec{r}_i$''.
	\end{itemize}
	Using the above declaration, in general the delta function is written as follows:
\begin{equation}
	\delta^3(\vec{r}-\vec r_i) = \left\{ \,
	\begin{IEEEeqnarraybox}[][c]{l?s}
		\IEEEstrut
		0 & $:\neg P$, \\
		\infty & $:P$.
		\IEEEstrut
	\end{IEEEeqnarraybox}
	\right.
	\label{deltaraw}
\end{equation}
This means that ``$\delta^3(\vec{r}-\vec r_i)$'' can be replaced by ``$0$'' when $\neg P$ \ul{and} by ``$\infty$'' when $P$, where each replacement is a process of computation. {\color{black}I may note that the definition of the delta function is only completed through the declaration of its behaviour under integration. However, here I am only concerned with how it behaves when it appears by itself to represent charge density in the first Maxwell equation, as known in the standard literature.}

Now, for the present purpose it is necessary to understand the logical relationship between these two processes of computation i.e. to figure out which {\it logical connective} (LC) relates these two processes of computation if delta function is written  in the following form:
\begin{eqnarray}
	\delta^3(\vec{r}-\vec r_i)=~[0:\neg P]~\text{LC}~ [\infty :P].\label{deltalc}
\end{eqnarray}
 At first sight it may appear that, both the truths must be valid in the process of construction of the logical form of the delta function, as I have already written the ``\ul{and}'' after (\ref{deltaraw}); therefore, ``LC'' must be replaced by ``$\wedge$ (AND)'' in (\ref{deltalc}). However, such a conception of the scenario should be avoided.	The ``\ul{and}'', written after (\ref{deltaraw}), is not a logical connective, but a part of the language (English) that is being used here to construct the logical structure of the computation process. This can be likened to Hilbert-Ackermann's declaration regarding the use of ``eq.'' in explaining logical connectives, in a footnote on p.6 of ref.\cite{hilbertacker}: ``{\it It should be noted that the abbreviation ``eq.'' used here does not belong to our set of logical symbols.}'', where the abbreviation stood for ``equivalent to''. In a more technical jargon, the ``\ul{and}'' written after (\ref{deltalc}) is part of the metalanguage English that is being used to construct the object language of logic that is written in terms of logical connectives e.g. see refs.\cite{fraenkel,reichenbach,tarskiundef} for related discussions. Indeed, consideration of the ``\ul{and}'', written after (\ref{deltaraw}), as a logical connective (i.e. ``$\wedge$(AND)''), leads to the truth of ``$(P\wedge\neg P)$'' which, in turn, leads to the falsehood of ``$(P\veebar\neg P)$'' i.e. violation of the LCL. The logical truth of the delta function abides by the LCL if and only if ``LC'' is replaced by ``$\veebar$ (EITHER...OR...)''  in (\ref{deltalc}). The situation can be explained by writing the following two propositions:
 \begin{itemize}
\item {\bf Q1}: $\delta^3(\vec{r}-\vec r_i)=~[0: \neg P]~\wedge~ [\infty : P],$
\item {\bf Q2}: $\delta^3(\vec{r}-\vec r_i)=~[0: \neg P]~\veebar~ [\infty : P],$
 \end{itemize}
 and then by writing the following truth table to analyze which one among {\bf Q1} and {\bf Q2} abides by the LCL.  
$$\begin{array}{|c|c|c|c|c|c|c|}\hline \delta^3(\vec{r}-\vec r_i)=0 : \neg P&\neg P & \delta^3(\vec{r}-\vec r_i)= \infty : P &P& P\veebar \neg P~ &\textbf{Q1}&\textbf{Q2}  \\\hline  F&F & T&T & T & F &T \\T&T & F&F &T&F&T\\ T&T & T&T &F&T&F\\F&F & F&F &F&T&F \\\hline\end{array}$$
It becomes clear that the truth values of {\bf Q2} only match with that of $P\veebar\neg P$ (LCL). So, abiding by the LCL, I write the delta function formally as {\bf Q2} i.e. 
\begin{eqnarray}
	\delta^3(\vec{r}-\vec r_i)=~[0:\neg P]~\veebar~ [\infty : P].\label{delta}
\end{eqnarray}
{\bf Digression:} Before proceeding further, I may worthily digress to mention that I have made two separate columns in the above truth table for ``$P$'' and ``$\neg P$'' to emphasize the equivalence between ``$\delta^3(\vec{r}-\vec r_i)=0$'' and ``$\vec{r}\neq \vec{r}_i$'', and between ``$\delta^3(\vec{r}-\vec r_i)=\infty$'' and ``$\vec{r}= \vec{r}_i$'', which are justified by the following two truth tables, from left to right, respectively. 
	$$\begin{array}{|c|c|c|}\hline \delta^3(\vec{r}-\vec r_i)=0&\vec{r}\neq\vec r_i ~ (\neg P) & \text{Validity}  \\\hline  F&F & T \\T&T & T \\ F&T & F \\T&F & F 	\\\hline\end{array}
	\qquad
	\begin{array}{|c|c|c|}\hline \delta^3(\vec{r}-\vec r_i)=\infty&\vec{r}=\vec r_i ~ (P) & \text{Validity}\\\hline  F&F & T \\T&T & T \\ F&T & F \\T&F & F 	\\\hline\end{array}
	$$
This, further, provides clarity regarding the sense in which the symbol ``$:$'' works in the present context, if one tries to make sense of it in terms of logical connectives. The above two truth tables manifest that the symbol ``$:$'' actually works as ``equivalence''/``if and only if'', i.e. ``$\equiv$'', in the context of the delta function while one tries to figure out its logical essence by looking at the $T$-s in the ``Validity'' columns of the above truth tables. In case it {\it were} ``$\delta^3(\vec{r}-\vec{r}_i)=\infty ~\text{IF}~ P$'' (i.e. ``$:$'' {\it were} replaced by ``$\leftarrow$'' to write ``$\delta^3(\vec{r}-\vec{r}_i)=\infty \leftarrow P$'') and ``$\delta^3(\vec{r}-\vec{r}_i)=0 ~\text{IF}~ \neg P$'' (i.e. ``$:$'' {\it were} replaced by ``$\leftarrow$'' to write ``$\delta^3(\vec{r}-\vec{r}_i)=0 \leftarrow \neg P$''), then in the last row of both the above truth tables, in the ``Validity'' column, ``$F$'' would have been replaced by ``$T$''.
}

{\color{black}
\section{Role of choice in human reasoning to write the first Maxwell's equation}\label{sec5}
	Having revisited the basic tools of logic and having written the delta function in terms of the logical connectives, I shall demonstrate the role of choice in human reasoning to write the first Maxwell's equation for a source point charge, a source configuration of point charges and a continuous distribution of source charge in the following subsections respectively.

\subsection{A source point charge} 
 I recast (\ref{divEdelta}) as follows:
	\begin{eqnarray}
		\vec{\nabla}\cdot\vec{E}_{(q_i, \vec{r}_i)}(\vec{r})
		&=& \frac{q_i}{\epsilon_0}\delta^3(\vec{r}-\vec r_i)~: \neg P\label{divEp},
	\end{eqnarray} 
where $\neg P$ is the condition for defining ``$\vec{E}_{(q_i, \vec{r}_i)}(\vec{r})$''. 
Using (\ref{delta}) in (\ref{divEp}), I obtain the following:
\begin{eqnarray}
\vec{\nabla}\cdot\vec{E}_{(q_i, \vec{r}_i)}(\vec{r})&=& \frac{q_i}{\epsilon_0}\[[0 :\neg P]~\veebar~ [\infty :P]\]: \neg P.\label{divEdecide}
\end{eqnarray}
Now, I need to figure out how ``$\neg P$'', sitting outside the bracket in (\ref{divEdecide}), affects the 
conditions inside the bracket in (\ref{divEdecide}). Since $\neg P$ (outside the bracket) must hold simultaneously alongside the rest within the bracket, I compute the validity conditions separately as follows:
\begin{eqnarray}
(\neg P\veebar P)\wedge (\neg P) ~\equiv~ (\neg P\wedge \neg P)\veebar (P\wedge \neg P)~\equiv~\neg P\veebar(P\wedge \neg P).\label{vcresult}
\end{eqnarray}
Certainly, as $P\wedge\neg P\equiv F$, the result of the above computation is $\neg P\veebar  F \equiv \neg P$. I {\it postpone} this step of computation (that eventually leads to (\ref{unprov})), because I intend to manifest the role of choices in human reasoning to write down the first Maxwell's equation as it appears in the standard literature. So, using (\ref{vcresult}) to intuitively compute the associated validity conditions corresponding to both the terms of (\ref{divEdecide}), I write (\ref{divEdecide}) as follows:
}
\begin{eqnarray}
\vec{\nabla}\cdot\vec{E}_{(q_i, \vec{r}_i)}(\vec{r})
	&=& [0: \neg P]~\veebar~ [\text{undecidable} : \underbrace{P\wedge \neg P}_{\text{\color{black}halt by choice}}].\label{divE}
\end{eqnarray}
{\color{black}This means I do not write ``$F$'' in place of ``$P\wedge \neg P$'', by choice, to create a decision problem.} The second term is undecidable due to the association of two contradictory conditions i.e. I can not decide what mathematical result I can write logically for the second term. Therefore, the computation halts. 

{\color{black} I may emphasize that obviously this decision problem is created due to the choice of postponement that has halted the computation. However, this choice should not be judged as right or wrong. Rather, it should be viewed as {\it the} human process of reasoning that makes room for further choices to yield the first Maxwell's equation as follows.}   
\begin{itemize}
\item {\bf Choice 1:}	I ignore ``$\neg P$'' in the second term and write the following:
\begin{eqnarray}
	\vec{\nabla}\cdot\vec{E}_{(q_i, \vec{r}_i)}(\vec{r})
	&=& [0:\neg P]~\veebar~ [\infty : P\underbrace{\wedge \neg P}_{\text{ignore by choice}}]\nonumber\\
	&=& [0: \neg P]~\veebar~ [\infty :P]~~~(\text{written by choice}).\label{choice1a}
\end{eqnarray}
``$\infty$'' is a result of a choice of ignorance -- a choice by which I ignore a necessary definability condition for ``$\vec{E}_{(q_i, \vec{r}_i)}(\vec{r})$'' on the left hand side. Thus, defying logic by invoking such a choice, {\color{black}using (\ref{delta}) in (\ref{choice1a})}, now I can write
\begin{eqnarray}
\vec{\nabla}\cdot\vec{E}_{(q_i, \vec{r}_i)}(\vec{r})=\frac{q_i}{\epsilon_0}\delta^3(\vec{r}-\vec{r}_i),\label{choice1}
\end{eqnarray}
 where the symbol ``$\vec{E}_{(q_i, \vec{r}_i)}(\vec{r})$'' is {\it undefined}. I may emphasize that  (\ref{choice1}) and (\ref{divEdelta}) should not be considered to be the same expression. It is only (\ref{choice1}) that can be found in standard literature and not (\ref{divEdelta}).
\item {\bf Choice 2:} I ignore $P$ in the second term and write the following:
\begin{eqnarray}
	\vec{\nabla}\cdot\vec{E}_{(q_i, \vec{r}_i)}(\vec{r})
	&=& [0: \neg P]~\veebar~ [0 :\underbrace{P\wedge}_{\text{ignore by choice}}\neg P]\nonumber\\
	&=& [0:\neg P]~\veebar~ [0 : \neg P].~~~(\text{written by choice})	\label{choice2}
\end{eqnarray}
Such a choice does not yield a new mathematical result in the second term other than ``$0$'' which is the result in the first term. Now, using the fact that $\neg P\veebar \neg P\equiv \neg P$ for any proposition $P$, I may write
\begin{eqnarray}
	\vec{\nabla}\cdot\vec{E}_{(q_i, \vec{r}_i)}(\vec{r})
	= 0~:\neg P.
\end{eqnarray}
\end{itemize}
Now, {\color{black}having explained the choices, I may perform the computation that I {\it postponed} a few paragraphs earlier.  Thus, I may simplify (\ref{vcresult}) as follows:
\begin{eqnarray}
	(\neg P\veebar P)\wedge (\neg P) ~\equiv~\neg P\veebar(P\wedge \neg P)~\equiv~ \neg P\veebar F~\equiv~\neg P,\label{p}
\end{eqnarray}	
which I intuitively use to write (\ref{divE}) as follows: 
\begin{eqnarray}
	&&\vec{\nabla}\cdot\vec{E}_{(q_i, \vec{r}_i)}(\vec{r})\nonumber\\
	&=& [0~:\neg P]\veebar[\text{undecidable}: F]~~~~(\text{\footnotesize corresponds to the last but one step}~\text{``}\neg P\veebar F\text{''} ~\text{\footnotesize in (\ref{p})})\nonumber\\
	&=& [0~:\neg P].~~~~~~~~~(\text{\footnotesize corresponds to the last step}~\text{``}\neg P\text{''} ~\text{\footnotesize in (\ref{p})})\label{divElogic}
\end{eqnarray}
} I may note that {\bf Choice 2} actually leads to the result that I can compute by following the usual rules of logic. Hence, two different cases arise from (\ref{divEdelta}). Either I illogically write (\ref{choice1}) by {\bf Choice 1}, or I logically write  (\ref{divElogic}) without making any choice. Certainly, which option one accepts is also a choice itself.  So, {\color{black}writing the meaning of ``$\neg P$'' explicitly,} I may conclude that the only logical answer is the following:
\begin{eqnarray}
	\vec{\nabla}\cdot\vec{E}_{(q_i, \vec{r}_i)}(\vec{r}) =0~:\vec{r}\neq\vec r_i.
\end{eqnarray}

\subsection{A source configuration of point charges}\label{sec5.1}
Now, I can extrapolate such a logical analysis for the electric field due to a configuration of non-overlapping point charges, although I admit that I am unable to take into account in symbolic terms that no two or more of the source point charges overlap. ``Non-overlapping'' is only a symbolically unexpressed assumption. Electric field at the field point $\vec{r}$, due to a source configuration of point charges $\{q_i\}$ situated at the points $\{\vec{r}_i\}$ such that $i\in[1,n]$, by the principle of superposition, is written as follows:
\begin{eqnarray}
\vec{E}_{\bigwedge_{i=1}^{n} (q_i,\vec{r}_i)}(\vec{r})=\sum_{i=1}^{n}\vec{E}_{(q_i,\vec{r}_i)}&=&\sum_{i=1}^{n}\left[ \frac{q_i}{4\pi\epsilon_0}\frac{(\vec{r}-\vec r_i)}{|\vec r-\vec r_i|^3}~:\vec{r}\neq\vec r_i \right]\nonumber\\
&=&\left[\sum_{i=1}^{n} \frac{q_i}{4\pi\epsilon_0}\frac{(\vec{r}-\vec r_i)}{|\vec r-\vec r_i|^3} \right]~:\vec{r}\neq\vec r_i~\forall~ i\in[1,n]. 
\end{eqnarray}
Here, the lower index ``$\bigwedge_{i=1}^{n} (q_i,\vec{r}_i)$'' stands for the phrase ``due to a source configuration of point charges $\{q_i\}$ situated at the points $\{\vec{r}_i\}$ such that $i\in[1,n]$'' i.e.  ``$\bigwedge_{i=1}^{n} (q_i,\vec{r}_i)$'' is the shorthand for ``$(q_1,\vec{r}_1)\wedge(q_2,\vec{r}_2)\wedge\cdots\wedge(q_n,\vec{r}_n)$''.
 Now, I write
\begin{eqnarray}
\vec{\nabla}\cdot\vec{E}_{\bigwedge_{i=1}^{n} (q_i,\vec{r}_i)}(\vec{r})=\sum_{i=1}^{n}\vec{\nabla}\cdot\vec{E}_{(q_i,\vec{r}_i)}&=&\left[\sum_{i=1}^{n} \frac{q_i}{4\pi\epsilon_0}\vec{\nabla}\cdot\frac{(\vec{r}-\vec r_i)}{|\vec r-\vec r_i|^3}\right]~:\vec{r}\neq\vec r_i~\forall~ i\in[1,n]\nonumber\\
&=&\left[ \sum_{i=1}^{n}\frac{q_i}{\epsilon_0}\delta^3(\vec{r}-\vec r_i)\right] ~:\vec{r}\neq\vec r_i~\forall~ i\in[1,n]\nonumber\\
&=&{\color{black}[[0 : Q]\veebar [\infty: \neg Q]]:Q},\label{divEconfigdelta}
\end{eqnarray}
{\color{black}where ``$Q$'' and ``$\neg Q$'' symbolize ``$\left\{\vec{r}=\vec r_i\text{ for some }i\in[1,n] \right\}$'' and ``$\left\{\vec{r}\neq\vec r_i~\forall~ i\in[1,n]\right\}$'' respectively, using which I have written 
	\begin{eqnarray}
		\sum_{i=1}^{n}\frac{q_i}{\epsilon_0}\delta^3(\vec{r}-\vec r_i)=[0 : Q]\veebar [\infty: \neg Q],\label{deltasum}
	\end{eqnarray}	
to write the final step of (\refeq{divEconfigdelta}). Then, following the course of reasoning similar to that for a source point charge, discussed previously, I choose to halt and create a decision problem by not writing ``$F$'' in place of ``$Q\wedge\neg Q$'' in the second term of (\refeq{divEconfigdelta}) and write the following:
}
 \begin{eqnarray}
 	\vec{\nabla}\cdot\vec{E}_{\bigwedge_{i=1}^{n} (q_i,\vec{r}_i)}(\vec{r})={\color{black}[0: Q]\veebar[\text{undecidable}: \underbrace{Q\wedge\neg Q}_{\text{halt by choice}}]}
 	\label{divEconfigdeltaformal}
 \end{eqnarray}
Here, I can make the following two choices.
\begin{itemize}
\item {\bf Choice 1:} I ignore ``$\neg Q$'' in the second term and write the following:
 \begin{eqnarray}
	\vec{\nabla}\cdot\vec{E}_{\bigwedge_{i=1}^{n} (q_i,\vec{r}_i)}(\vec{r})
	&=&\left[0:\neg Q\right]~\veebar~\left[\infty :Q\underbrace{\wedge\neg Q}_{\text{ignore by choice}}\right]~~~~~~\nonumber\\
	&=&\left[0: \neg Q\right]~\veebar~\left[\infty :Q\right]\qquad\qquad\text{(written by choice)}.
\end{eqnarray}
``$\infty$'' is a result of a choice of ignorance -- a choice by which I ignore a necessary definability condition for one of the ``$\vec{E}_{(q_i, \vec{r}_i)}(\vec{r})$''-s which is a part of the construction of ``$\vec{E}_{\bigwedge_{i=1}^{n} (q_i,\vec{r}_i)}(\vec{r})$'' on the left hand side. Thus, defying logic by invoking such a choice, now I can write
\begin{eqnarray}
	\vec{\nabla}\cdot\vec{E}_{\bigwedge_{i=1}^{n} (q_i,\vec{r}_i)}(\vec{r})=\sum_{i=1}^{n}\frac{q_i}{\epsilon_0}\delta^3(\vec{r}-\vec{r}_i)\label{configchoice1}
\end{eqnarray}
by using (\ref{delta}), where the symbol ``$\vec{E}_{\bigwedge_{i=1}^{n} (q_i,\vec{r}_i)}(\vec{r})$'' is {\it undefined}. I may emphasize that  (\ref{configchoice1}) and (\ref{divEconfigdelta}) should not be considered to be the same expression. It is only (\ref{configchoice1}) that can be found in standard literature and not (\ref{divEconfigdelta}).

\item {\bf Choice 2:} I ignore ``$Q$'' in the second term and write the following:
 \begin{eqnarray}
	\vec{\nabla}\cdot\vec{E}_{\bigwedge_{i=1}^{n} (q_i,\vec{r}_i)}(\vec{r})
		&=&\left[0:\neg Q\right]~\veebar~\left[0 :\underbrace{Q\wedge}_{\text{ignore by choice}}\neg Q\right]\nonumber\\
	&=&\left[0:\neg Q\right]~\veebar~\left[0 :\neg Q\right]\qquad\qquad\text{(written by choice)}.
\end{eqnarray}
Such a choice does not yield a new mathematical result in the second term other than ``$0$'' which is the result in the first term. Now, using the fact that $\neg Q\veebar \neg Q\equiv \neg Q$, I may write
 \begin{eqnarray}
	\vec{\nabla}\cdot\vec{E}_{\bigwedge_{i=1}^{n} (q_i,\vec{r}_i)}(\vec{r})= 0:\neg Q.
\end{eqnarray}
\end{itemize}
Now, I strictly adhere to the rules of logic, {\color{black}without choosing to halt, and proceed from (\ref{divEconfigdeltaformal}) as follows similar to the case of the source point charge discussed previously. 
\begin{eqnarray}
	\vec{\nabla}\cdot\vec{E}_{\bigwedge_{i=1}^{n} (q_i,\vec{r}_i)}(\vec{r})
	&=&\left[0:\neg Q\right]~\veebar~\left[\text{undecidable}: Q\wedge\neg Q\right]\nonumber\\
	&=&\left[0:\neg Q\right]~\veebar~\left[\text{undecidable}: F\right]\qquad{\footnotesize (\text{using}~ \neg Q\wedge Q\equiv F )}\nonumber\\
    &=&\left[0: \neg Q \right]\qquad{\footnotesize (\text{using}~ \neg Q\veebar F\equiv \neg Q )}.\label{divEconfiglogic}
\end{eqnarray}}
I note that {\bf Choice 2} actually leads to the result that I can achieve just by following the usual rules of logic. Hence, two different cases arise from (\ref{divEconfigdelta}). Either I illogically write (\ref{configchoice1}) by {\bf Choice 1}, or I logically write  (\ref{divEconfiglogic}) without making any choice. Certainly, which option one accepts is also a choice itself. However, {\color{black} writing the meaning of ``$\neg Q$'' in (\ref{divEconfiglogic}) explicitly, I conclude that} the only logical answer is the following:
 \begin{eqnarray}
 	\vec{\nabla}\cdot\vec{E}_{\bigwedge_{i=1}^{n} (q_i,\vec{r}_i)}(\vec{r})=0~:\vec{r}\neq\vec r_i~\forall~ i\in[1,n].
\end{eqnarray}

 \subsection{A continuous distribution of source charge}\label{sec5.2}
 The calculation of the electric field due to a continuous distribution of source charge consists of the following steps, which can be found in the standard modern textbooks\cite{jackson,griffiths}, albeit devoid of the symbolic precision that is being discussed here. At first, I consider an infinitesimal volume element $d\tau^{\prime}$ that contains infinitesimal source charge $dq=\rho(\vec{r}^{~\prime})d\tau^{\prime}$. Then, I write the electric field at the field point $\vec{r}$ due to the infinitesimal source charge $dq$, situated at $\vec{r}^{~\prime}$, by replacing ``a point charge'' by ``an infinitesimal charge''  
 in (\ref{epc}), as follows:
 \begin{eqnarray}
 	\vec{E}_{(dq, \vec{r}^{\prime})}(\vec{r})&=&\frac{dq}{4\pi\epsilon_0} \frac{(\vec{r}-\vec{r}^{~\prime})}{|\vec{r}-\vec{r}^{~\prime}|^3}~~~:\vec{r}\neq\vec{r}^{~\prime}
 	\nonumber\\
\equiv~\vec{E}_{(\rho(\vec{r}^{\prime})d\tau^{\prime}, \vec{r}^{\prime})}(\vec{r}) &=& \frac{\rho(\vec{r}^{~\prime})d\tau^{~\prime}}{4\pi\epsilon_0} \frac{(\vec{r}-\vec{r}^{~\prime})}{|\vec{r}-\vec{r}^{~\prime}|^3}\quad:\vec{r}\neq\vec{r}^{~\prime}~~\text{ [using $dq=\rho(\vec{r}^{~\prime})d\tau^{\prime}$]}.
 \end{eqnarray}
Then, I write the electric field at the field point $\vec{r}$ due to a continuous distribution of source charge $\rho(\vec{r}^{~\prime})$ in a volume $\tau^{\prime}$, as follows:
\begin{eqnarray}
 \vec{E}_{(\rho(\vec{r}^{\prime}), \tau^{\prime})}(\vec{r})=\frac{1}{4\pi\epsilon_0}\int_{\tau^{\prime}} \frac{\rho(\vec{r}^{~\prime})(\vec{r}-\vec{r}^{~\prime})}{|\vec{r}-\vec{r}^{~\prime}|^3}~d\tau'\quad:\vec{r}\not\in\tau^{\prime}.
\end{eqnarray}
The subscript ``$(\rho(\vec{r}^{\prime}), \tau^{\prime})$'' stands for the phrase ``due to a continuous distribution of source charge $\rho(\vec{r}^{~\prime})$ in a volume $\tau^{\prime}$''. The condition ``$\vec{r}\not\in\tau^{\prime}$'' means ``the field point $\vec{r}$ does not lie within the volume $\tau^{\prime}$.''

Here, I may admit that, unlike the symbolically logical passage from the case of a point source charge ``$\vec{E}_{(q_i, \vec{r}_i)}(\vec{r})$'' to the case of a collection of non-overlapping point source charges ``$\vec{E}_{\bigwedge_{i=1}^{n} (q_i,\vec{r}_i)}(\vec{r})$'', I can not find a symbolically logical passage from the case of an infinitesimal source charge ``$\vec{E}_{(\rho(\vec{r}^{\prime})d\tau^{\prime}, \vec{r}^{\prime})}(\vec{r})$'' to the case of a continuous distribution of source charge ``$\vec{E}_{(\rho(\vec{r}^{\prime}), \tau^{\prime})}(\vec{r})$''. {\color{black}This goes down to the fact that an infinitesimal charge, in an infinitesimal volume element, is neither a point charge nor not a point charge; so, it can be thought of as either, according to need e.g. see Appendix (\ref{appB}) to see how this fact is used to construct the continuity equation, which is an elementary example of a decision problem.} Therefore, the passage from 
``$\vec{E}_{(\rho(\vec{r}^{\prime})d\tau^{\prime}, \vec{r}^{\prime})}(\vec{r})$'' to ``$\vec{E}_{(\rho(\vec{r}^{\prime}), \tau^{\prime})}(\vec{r})$'' is rather symbolically {\it intuitive}.

Now, using the standard steps of calculation\cite{jackson,griffiths}, I can write the following:
\begin{eqnarray}
\vec{\nabla}\cdot\vec{E}_{(\rho(\vec{r}^{\prime}), \tau^{\prime})}(\vec{r})&=&\frac{1}{4\pi\epsilon_0}\int_{\tau^{\prime}}\rho(\vec{r}^{~\prime}) \vec{\nabla}\cdot\frac{(\vec{r}-\vec{r}^{~\prime})}{|\vec{r}-\vec{r}^{~\prime}|^3}~d\tau'\quad:\vec{r}\not\in\tau^{\prime}\nonumber\\
&=&\frac{1}{\epsilon_0}\int_{\tau^{\prime}}\rho(\vec{r}^{~\prime}) \delta^3(\vec{r}-\vec{r}^{~\prime})~d\tau'\quad:\vec{r}\not\in\tau^{\prime}~~\left[\because \vec\nabla\cdot\frac{(\vec{r}-\vec{r}^{~\prime})}{|\vec{r}-\vec{r}^{~\prime}|^3}=4\pi\delta^3(\vec{r}-\vec{r}^{~\prime})\right].\nonumber\\
&&\hfill ~~\label{econt}
\end{eqnarray}
Further, using the properties of the delta function\cite{dirac,jackson,griffiths}, I may write the following formal structure:
\begin{eqnarray}
\int_{\tau^{\prime}}\rho(\vec{r}^{~\prime}) \delta^3(\vec{r}-\vec{r}^{~\prime})~d\tau'&=&~[0:\vec{r}\not\in\tau^{\prime}]~\veebar~[\rho(\vec{r}):\vec{r}\in\tau^{\prime}]\nonumber\\
&=& {\color{black} ~[0:\neg R]~\veebar~[\rho(\vec{r}):R]},\label{deltaformal}
\end{eqnarray}
{\color{black} where I have symbolized ``$\vec{r}\in\tau^{\prime}$'' and ``$\vec{r}\not\in\tau^{\prime}$'' as ``$R$'' and ``$\neg R$'', respectively.} Therefore, using (\ref{deltaformal}) {\color{black} and following the steps of reasoning similar to the case of a source point charge discussed previously}, I may recast (\ref{econt}) in the following way: 
\begin{eqnarray}
\vec{\nabla}\cdot\vec{E}_{(\rho(\vec{r}^{\prime}), \tau^{\prime})}(\vec{r})&=&[[0:\neg R]~\veebar~[\rho(\vec r): R]]:\neg R\nonumber\\
&=&[0:\neg R]~\veebar~[\text{undecidable }: \underbrace{R\wedge\neg R}_{\text{\color{black}halt by choice}}]\nonumber\\
\label{econtdelform}.
\end{eqnarray}
The second term is undecidable because of the contradictory conditions i.e. I have arrived at a decision problem. Now, I can make two choices like in the previous scenarios. 
\begin{itemize}
\item {\bf Choice 1:} I ignore ``$\neg R$'' in the second term and write the following:
\begin{eqnarray}
\vec{\nabla}\cdot\vec{E}_{(\rho(\vec{r}^{\prime}), \tau^{\prime})}(\vec{r})
&=&[0:\neg R]~\veebar~[\rho(\vec r)/\epsilon_0:R\underbrace{\wedge \neg R}_{\text{ignore by choice}}]\nonumber\\
&=&[0:\neg R]~\veebar~[\rho(\vec r)/\epsilon_0: R]\qquad\qquad\text{(written by choice)}\label{contchoice1}.
\end{eqnarray} 
``$\rho(\vec{r})$'' is a result of a choice of ignorance -- a choice by which I ignore the definability condition of ``$\vec{E}_{(\rho(\vec{r}^{\prime}), \tau^{\prime})}(\vec{r})$'' on the left hand side. Thus, it is only by defying logic, I am able to write some non-trivial result, namely $\rho(\vec{r})$, other than ``$0$'' in the second term.
\item {\bf Choice 2:} I ignore ``$R$'' in the second term and write the following:
\begin{eqnarray}
	\vec{\nabla}\cdot\vec{E}_{(\rho(\vec{r}^{\prime}), \tau^{\prime})}(\vec{r})
	&=&[0: \neg R]~\veebar~[0: \underbrace{R\wedge}_{\text{ignore by choice}}\neg R]\nonumber\\
	&=&[0:\neg R]~\veebar~[0:\neg R]~~\text{(written by choice)}.
\end{eqnarray} 
Such a choice does not yield a new mathematical result in the second term other than ``$0$'' which is the result in the first term. Now, using the fact that $\neg R\veebar \neg R\equiv \neg R$, I can write
\begin{eqnarray}
	\vec{\nabla}\cdot\vec{E}_{(\rho(\vec{r}^{\prime}), \tau^{\prime})}(\vec{r})
	&=&0:\neg R.
\end{eqnarray} 
\end{itemize}

Now, I strictly adhere to the rules of logic and proceed {\color{black} from (\ref{econtdelform}), without choosing to halt and following the course of reasoning alike the case of a source point charge discussed previously,}  as follows.
\begin{eqnarray}
	\vec{\nabla}\cdot\vec{E}_{(\rho(\vec{r}^{\prime}), \tau^{\prime})}(\vec{r})
	&=&\left[0:\neg R\right]\veebar~[\text{undecidable}: R\wedge\neg R]\nonumber\\
	&=&\left[0:\neg R\right]\veebar~[\text{undecidable}: F] \qquad{\footnotesize (\text{using}~R\wedge\neg R\equiv F)}\nonumber\\
	&=&\left[0:\neg R\right]\qquad\footnotesize{(\text{using}~\neg R\veebar F\equiv \neg R)}.\label{divEcontlogic}
\end{eqnarray}
I note that {\bf Choice 2} actually leads to the result that I can achieve just by following the usual rules of logic. Hence, two different cases arise from (\ref{econtdelform}). Either I illogically write (\ref{contchoice1}) by {\bf Choice 1}, or I logically write  (\ref{divEcontlogic}) without making any choice. Certainly, which option one accepts is also a choice itself. However, the only logically valid answer is the following:
\begin{eqnarray}
	\vec{\nabla}\cdot\vec{E}_{(\rho(\vec{r}^{\prime}), \tau^{\prime})}(\vec{r})
	&=&0:\vec{r}\not\in\tau^{\prime},
\end{eqnarray} 
where I have explicitly written the meaning of ``$\neg R$''.

Thus, I may conclude that the first Maxwell's equation, symbolically identifiable as ``$\vec{\nabla}\cdot\vec{E}(\vec{r})=\rho(\vec{r})/\epsilon_0$'' from the standard modern texts\cite{jackson,griffiths} is formally unprovable in light of ECC. Since this equation is one of the founding premises of the standard ``proof'' of  Poynting's theorem available in the standard modern textbooks\cite{jackson, griffiths}, then Poynting's theorem is rendered formally unprovable as well. Consequently, the explanation of `light' as `propagation of electromagnetic energy' comes into question on theoretical ground. 


\section{Conclusion}\label{sec6}
In this work, following basic rules of logic, I have shown that the first Maxwell's equation, symbolically  identifiable as ``$\vec{\nabla}\cdot\vec{E}(\vec{r})=\rho(\vec{r})/\epsilon_0$'' from the standard modern texts\cite{jackson,griffiths}, is formally unprovable in light of EPR Completeness Condition\cite{epr}. Although EPR, themselves, only used such completeness condition to analyze certain consequences in quantum mechanics\cite{epr}, but the present analysis concerning the basics of classical electrodynamics only brings out a broader aspect of such completeness condition that underlies the logic and language of physics in general, irrespective of `classical' and `quantum'. To put the emphasis along similar matters of fact, I may further point out that in contrast to the very recent trend of relating decision problems uniquely to quantum  physics\cite{udgap1,udgap2,udgap3,udqua1,udqua2}, the present work only shows that decision problems have no unique connection with quantum physics -- rather it is a part of human reasoning and language with which physics is done {\color{black} (see Appendix (\ref{appB}) as well)}. 

 Last but not the least, there is a very different issue that is associated with the conclusion of this work. Due to the unprovability of the Poynting's theorem, now the explanation of `light' as `propagation of electromagnetic energy' is jeopardized from the theoretical point of view. This, in turn, provides the room and the motivation to rethink about Tesla's objections to Maxwell-Hertz theory\cite{tesla}. {\color{black} While this may appear, at first, to be only of historical significance, a bit of careful thinking reveals some deeper facts. It must be remembered that light can not be seen to propagate either as wave or particle (photon). Propagation of light in itself is only a theoretical fact. Wave model is based on Maxwell's equations\cite{maxwelleq,poynting} and photon model is often constructed in terms of wave function\cite{qoptics2008}. Such theoretical foundations let us interpret experiments\cite{smith2007,aspect2007,centphoton,nobel2022}. Loss of such theoretical foundations provides us a new opportunity to think about the interpretations of optical experiments, especially those involving entangled photons which epitomize the success of modern science\cite{nobel2022}.} I intend to report further along such lines of investigation in future.

\vspace{0.3cm}

{\it Acknowledgment:} This work has been supported by  the Department of Science and Technology of India through the  INSPIRE Faculty Fellowship, Grant no.- IFA18-PH208. The author thanks an anonymous referee for the insightful remarks that resulted in the refinement of this manuscript to reach its present version.

{\it Declaration:} On behalf of all authors, the corresponding author states that there is no conflict of interest.

\appendix
\section{An estimate of $s, s'$: computational content of Maxwell's statement of Coulomb's hypothesis}\label{appA}
Maxwell's statement of Coulomb's hypothesis, in light of ECC, calls for an explanation of the length scales $s$ and $s'$, that characterize the electrified bodies, in (\ref{maxclss}) which must appear only as an approximate expression along with correction terms in powers of $s/r, s'/r$, etc. such that these terms become negligible as $r\ggg s, s'$ in tandem with the experimental data that verifies Coulomb's hypothesis. It should appear as if the electrified bodies themselves are the most natural uncertainties, characterized by $s, s'$, in the experimental demonstration of the Coulomb's hypothesis. Such an analysis indeed has been penned for gravitational interaction between two bodies in ref.\cite{majhidot}, which respects experimental observations. While I plan to report the details of such analysis for the interaction between two electrified/charged bodies on a different occasion, here I shall provide a glimpse of the relevant outcome of the analysis which will show an estimate of $s, s'$ in (\ref{maxclss}). In what follows, to make it familiar in accord with the modern notations available in the textbooks, I shall write $q_1, q_2$ in stead of $e, e'$. Also, for similarity of notation, I shall write $\xi_1, \xi_2$ in stead of $s, s'$.

The analysis that has been presented in ref.\cite{majhidot}, for the case of gravity, takes into account the extension of the dot that is needed to demonstrate the notion of ``a point'' and hence its smallness can only be understood in relation to what we may call ``a line''. Adopting such a philosophy of incorporating the natural uncertainties that are necessary for demonstration of numbers by making cuts on a line, and following the steps of reasoning presented in ref.\cite{majhidot}, the relevant expressions for the present discussion, from which the analysis shall begin, can be written as follows:
\begin{eqnarray}
	\frac{F}{F_C}=\frac{\xi_1\xi_2}{d^2}~~:~F_C=\frac{m^2c^3}{h},~\xi_i=\zeta\frac{q_i}{m}~~\forall i\in[1,2],~\zeta=\left(\frac{h}{4\pi\epsilon_0c^3}\right)^{\frac{1}{2}}.\label{cons}
\end{eqnarray}
In comparison to the notations of ref.\cite{majhidot}, and in order to manifest the difference between the quantities involved, I have written $F_C, \xi_1, \xi_2$ here in place of $F_0, s_1, s_2$ from ref.\cite{majhidot}, respectively. Here, 
\begin{eqnarray}
	\epsilon_0&=&	8.854\times 10^{-12}~\text{C}^2.\text{kg}^{-1}.\text{meter}^{-3}.\text{sec}^2\\
	h&=&6.626\times 10^{-34}~\text{kg}.\text{meter}^2.\text{sec}^{-1}\\
	c&=&3\times 10^8~\text{meter}.\text{sec}^{-1}\\
	\therefore \zeta= \left(\frac{h}{4\pi\epsilon_0c^3}\right)^{\frac{1}{2}}&=&4.696\times 10^{-25}~\text{C}^{-1}.\text{kg}.\text{meter}.
\end{eqnarray}
For $m=1$ kg and $q_i=1$ C, $\xi_i=4.696\times 10^{-25}$ meter. 

By construction, $d=r_i+\xi_i\forall i\in[1,2]$. The condition $ d>2\xi_i \equiv r_i>\xi_i$ is called large distance w.r.t. $\xi_i$ and the condition $ d<2\xi_i \equiv r_i<\xi_i $ is called small distance w.r.t. $\xi_i$. 
 
For large distance analysis w.r.t $\xi_1$: 
\begin{eqnarray}
	\frac{F}{F_C}&=&\frac{\xi_2\xi_1}{r_1^2}\left[1-\frac{2\xi_1}{r_1}+\frac{3\xi_1^2}{r_1^2}-\cdots\right]~:~d>2\xi_1, ~r_1>\xi_1 \label{xi1}\\
	\equiv F&=&\frac{1}{4\pi\epsilon_0}\frac{q_2q_1}{r_1^2}\left[1-2\zeta\frac{q_1}{r_1}+3\zeta^2\frac{q_1^2}{r_1^2}-\cdots\right]~:~r_1>\zeta\frac{q_1}{m}\qquad[\text{using} (\ref{cons})].
\end{eqnarray}
For large distance analysis w.r.t $\xi_2$: 
\begin{eqnarray}
	\frac{F}{F_C}&=&\frac{\xi_1\xi_2}{r_2^2}\left[1-\frac{2\xi_2}{r_2}+\frac{3\xi_2^2}{r_2^2}-\cdots\right]~:~d>2\xi_2, ~r_2>\xi_2 \label{xi2}\\
	\equiv F&=&\frac{1}{4\pi\epsilon_0}\frac{q_1q_2}{r_2^2}\left[1-2\zeta\frac{q_2}{r_2}+3\zeta^2\frac{q_2^2}{r_2^2}-\cdots\right]~:~r_2>\zeta\frac{q_2}{m}\qquad[\text{using} (\ref{cons})].
\end{eqnarray}
The following may be called the point charge limit w.r.t. both $\xi_1$ and $\xi_2$:
\begin{eqnarray} 
	 \frac{F}{F_C}&\simeq& \frac{\xi_1\xi_2}{r^2}: d=r_i+\xi_i\simeq r_i\simeq r ~(\text{say}) \text{~when~} r_i\ggg\xi_i~\forall i\in[1,2],\label{pointcharge}\\
\equiv {F}&\simeq&\frac{1}{4\pi\epsilon_0} \frac{q_1q_2}{r^2}~:~
r\simeq r_i\ggg\xi_i~\forall i\in[1,2]\quad(\text{using (\ref{cons})}).
\end{eqnarray}
This provides a justification of Maxwell's statement of Coulomb's hypothesis along with an estimate of the length scales $\xi_1,\xi_2$ which characterize the two charged bodies. The essence of the associated condition, and why this can be called the point charge limit, can be understood by calling a point as that that is extremely small compared to what can be called a line -- a refinement of the first axiom of geometry by considering the concreteness of a dot that is necessary for the purpose of demonstration of what a point is\cite{majhidot}.

In passing I may note that the small distance analysis, w.r.t. both $\xi_1$ and $\xi_2$, is well behaved as follows:
\begin{eqnarray}
	F&=&F_C\left[1-2\frac{r}{\xi}+3\frac{r^2}{\xi^2}-\cdots\right]~:~\xi_1=\xi_2=\xi>r=r_1=r_2,\label{sd}\\
	\equiv F&=&\frac{m^2c^3}{h}\left[1-2\frac{r}{\zeta q}+3\frac{r^2}{\zeta^2q^2}-\cdots\right]~:~r<\zeta\frac{q}{m}\qquad(\text{using (\ref{cons})}).\label{sdq}
\end{eqnarray}
All the above expressions and the corresponding analysis can be cast into a form that manifests the role of fine structure constant. However, such discussions are not necessary for the present purpose and will be reported in future on a different occasion.


\section{A decision problem as the basis of the continuity equation}\label{appB}

\begin{figure}[hbt]\label{fig}
	\begin{center}
		\includegraphics[scale=0.40]{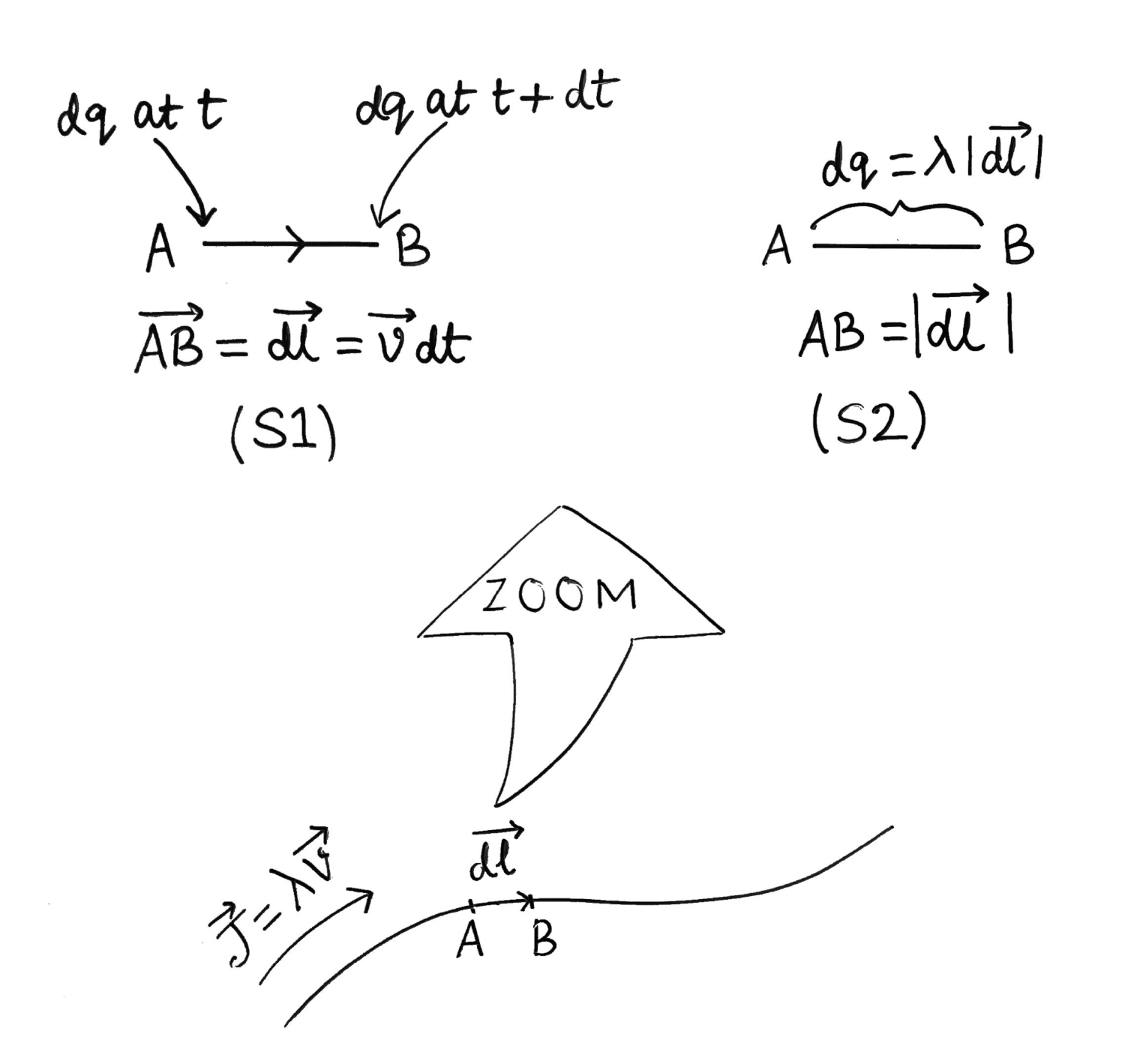}
	\end{center}
\end{figure}

The continuity equation, the so called local conservation law, forms the basis of our general understanding of flow of quantities like charge, mass, etc. in physics e.g. see refs.\cite{batchelor,marsden,landau,griffiths,jackson}. Here, I point out how the continuity equation is founded on a decision problem that allows for making suitable choices concerning two apparently contradictory propositions.  For the present discussion it is convenient to consider the continuity equation involving charge and current\cite{griffiths}. To understand the issue at hand, it is sufficient and convenient to discuss the flow of charge along a wire that can be written as follows:
\begin{eqnarray}
\vec
\nabla\cdot\vec J+\frac{\partial \lambda}{\partial t}=0,\label{con}
\end{eqnarray}
 where $\vec J=\lambda\vec v$, $\vec v=d\vec \ell/dt$, $d\vec{\ell}={dx}~\hat i+{dy}~\hat j+{dz}~\hat k$ and $\vec v$ satisfies the condition $\vec{\nabla}\cdot\vec v=0$\cite{landau, batchelor, marsden}; $\lambda[x(t), y(t), z(t), t]$ is the linear charge density (charge per unit length) at some instant $t$, $\vec J[x(t), y(t), z(t), t]$ is the line current density and $[x(t), y(t), z(t)]$ denote the spatial coordinate of any point on the line of flow at some time $t$. The construction of the relation ``$\vec J=\lambda \vec v$'' is based on two scenarios of interpreting the charge flow that can be demonstrated through figure (\ref{fig}), which can be generalized for surface and volume flows e.g. see ref.\cite{griffiths}. 
\begin{itemize}
\item {\bf Scenario 1 (S1):} Considering that a point charge $dq$ is displaced through $\vec{dl}$  in time $dt$ such that we can write  $\vec{dl}=\vec{v}dt$.  Then, we can write $|\vec{dl}|=|\vec{v}|dt$ and hence, $dq=\lambda |\vec{v}|dt$. In this case, the charge $dq$ is at the point $A$ at some time $t$ and then it is at the point $B$ at some time $t+dt$, so that the displacement in time duration $dt$ is $\vec{AB}=\vec{dl}$. In such a scenario the following assertion holds true:
\begin{eqnarray}
	M \equiv \text{$dq$ is a point charge}.
\end{eqnarray}

\item {\bf Scenario 2 (S2):} At any instant $t$ of the passage of a line current, there is a line charge density $\lambda$ (charge per unit length) such that the charge $dq$ distributed over the length $d\ell$ is written as $dq=\lambda dx$. In this case, the charge $dq$ is neither at point $A$, nor at point $B$, but it is spread over the line $AB$. Therefore, the following proposition holds true for such explanation:
\begin{eqnarray}
\neg M \equiv \text{$dq$ is NOT a point charge}.
\end{eqnarray}
\end{itemize}
Therefore, the relation $\vec{J}= dq/dt=\lambda\vec{v}$ holds true if and only if $M$ and $\neg M$ in one and the same process of reasoning that leads to such a construction. It may be written as
\begin{eqnarray}
\vec{J}=\lambda\vec{v}~: M\wedge\neg M. 
\end{eqnarray}
The situation can be viewed as a decision problem where it can not be decided whether $dq$ is a point charge or not\cite{hilbertacker}. This, however, allows for suitable choices to be made according to context i.e. $dq$ is considered as a point charge in {\bf S1} and it is not considered as a point charge in {\bf S2}.

\end{document}